# An Influence Diagram-Based Approach for Estimating Staff Training in Software Industry

Kawal Jeet,[1] Vijay Kumar Mago,[1] Bhanu Prasad[2] and Rajinder Singh Minhas[3]

[1]*Department of Computer Science, DAV College, Jalandhar, India;* [2]*Department of Computer and Information Sciences, Florida A&M University, Tallahassee, FL 32307, USA;* [3]*Department of Computer Science, MLUDAV College, Phagwara, India*

**ABSTRACT**

The successful completion of a software development process depends on the analytical capability and foresightedness of the project manager. For the project manager, the main intriguing task is to manage the risk factors as they adversely influence the completion deadline. One such key risk factor is staff training. The risk of this factor can be avoided by pre-judging the amount of training required by the staff. So, a procedure is required to help the project manager make this decision. This paper presents a system that uses influence diagrams to implement the risk model to aid decision making. The system also considers the cost of conducting the training, based on various risk factors such as, (i) Lack of experience with project software; (ii) Newly appointed staff; (iii) Staff not well versed with the required quality standards; and (iv) Lack of experience with project environment. The system provides estimated requirement details for staff training at the beginning of a software development project.

## 1. INTRODUCTION

Several factors pose a challenge to the success of Software Development Projects (SDPs) (Hui et al., 2004), for example, low productivity, creeping user

---

requirements, staff experience shortage, incapable project management and so on. These factors have an impact on the performance, schedule, and cost of the project (Boehma et al., 2000). Staff training is one of the key factors that adequately influence the SDP (Boehma, 1991). Subjective ways to find the impact of this factor on SDP are common in the software industry, but because of their opaque nature, are found to be inappropriate. For instance, the use of expert opinion in a traditional way, without any scientific theory or technology backup, is not sufficient in today's complex business environment. So, a more scientific approach is required to improve the success rate and outcome of the projects. In this paper, we have applied Influence Diagram (ID) (Uffe et al., 2008) as a scientific approach to the subjective technique of risk management to calculate the percentage of staff training required and the cost of conducting this training. A tool 'Management of Staff Training' (MaST) has been developed for this purpose. Some preliminary details of this tool are available in Jeet et al. (2008).

### 1.1 Risk Management

Despite all the efforts on the development methodologies and processes, many software projects are still unsuccessful, either failing to meet the goals, going over budget, finishing late, or finishing without a satisfied end-user. Some projects fail entirely or are canceled long before their delivery date. According to an independent survey conducted by an agency, only 26% of the projects are delivered on time, whereas 29% are completely canceled, and the rest happen to be late due to poor risk management (Risk Management: Web address). The reason for the high failure rate is that the project managers are not managing the circumstances or events (called risk or risk factor) that have potential to cause damage to the project, implying that the project managers are not applying risk management techniques. Risk Management is a collection of methods that can minimize the impact of risks on the project. Managing risk is a continuous process that has to be followed throughout the life cycle of the software development from the investigation of concept up till the delivery of the project (McManus, 2003).

Risk management is the process of applying appropriate tools and methods to keep the risk within tolerable limits (Charette, 1989). Risk management consists of several sub-activities such as:

- *Risk Identification*
- *Risk analysis*
- *Risk Planning*
- *Risk Tracking*
- *Risk Control*

The focus of our research is *risk prioritization*—a sub activity of risk analysis phases. Risk prioritization helps to focus on the project's most severe risks by calculating their exposure. 'Risk Exposure' is the product of the probability of occurrence of a risk and the potential magnitude of the loss that could be caused due to that risk. The details are given in Section 3.3.

## 1.2 Influence Diagrams

Influence Diagram (ID) also called 'Causal Network Modeling' is a highly acknowledged method for combining related probabilistic events into a single mathematical model (Park et al., 1998). The modeling is based on probabilistic theory. Influence Diagrams are ideally suitable for situations involving reasoning and decision making under uncertainty.

The network contains 'chance' nodes that represent the events and their probability of occurrence. Events may be independent or conditionally dependent on other events. A 'utility' node represents the desirability of different event combinations involved in the network. The nodes that are used in the construction of the ID for MaST are briefly described below:

1. *Chance nodes* are random variables drawn as circles or ovals. They represent uncertain events relevant to the decision problem.
2. *Value nodes* are drawn as diamonds and represent the utility, i.e., a measurement of outcomes of the decision process. They are quantified by each of the possible combinations of outcomes of the parent nodes.

We have used a software named GeNIe (GeNIe: Web Address) for the construction of ID.-Normally, an arc in an ID denotes an influence i.e., the node at the tail (A or B in Figure 1) of the arc influences the value (or the probability distribution over the possible values) of the node at the head (C in Figure 1) of the arc. For example, in Figure 1 the events A and B have an effect on the occurrence of C.

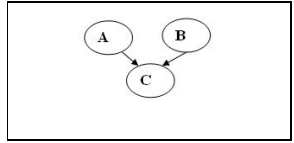

**Fig 1:** Cause-effect relationship

This is reflected by drawing the arcs from A to C and B to C as shown above. A and B are causes (c) and C is the effect (e). Now let us discuss the mathematics involved.

$$P(e|c) = \frac{P(c|e).P(e)}{P(c)}$$

This equation is called Baye's theorem (McCabe, 2001). Here, P(e|c) is the conditional probability of the effect C being true given the cause A or B, also known as the posterior probability or posterior. P(e) is the probability of the effect 'e' being true or the prior probability of effect. P(c|e) is the probability of occurring of cause 'c', given that the effect e is true. P(c) is the probability of the cause c being true or the prior probability of the cause.

ID provides the following contributions for modeling the management process:

1. An environment in which experts and decision makers can discuss the management of the problem and interdependencies of decisions and events, without invoking any formal mathematical, probabilistic or statistical notation. ID links the real-world decision making with that of a scientifically trained analyst. The decision-maker focuses on the problem and ensures that the ID accurately reflects the real world.
2. It helps reducing large volumes of data that is essential to the decision-making process.
3. It provides a sensitivity analysis to show how much effect the particular decisions or uncertain events have on the final outcomes.

## 2. REVIEW OF LITERATURE

Since the early 1980s, IDs have been used in a wide variety of applications like medical diagnosis (Mago et al., 2008), clinical decisions (Owens et al., 1997), cyber crime detection (Abouzakhar et al., 2003), etc. Risk management involves several factors, such as diverse processes and product variables, experiential evidence and expert judgment, cause and effect relationships, uncertainty, incomplete information, and so on. ID is best suited under such circumstances.

Bayesian Networks (BNs), a parent of ID, have been used for two software metrics, namely defect modeling/prediction and resource modeling/prediction (Norman et al., 2000). Some authors (Fan et al., 2004) developed a Bayesian Belief Network-based procedure to calculate the dynamic resource adjustment in the software industry. Similarly, researchers (Chee et al., 1995) used an ID to analyze the metrics data, collected at various stages of software development cycle, to identify the sources of errors, and to identify the future risk items like reliability, schedule, complexity or non-reusability. IDs have also been used to maintain project delays based on expert's opinion (Melo et al., 2007). Their main goal was to establish a relationship between project developers, software and any task that can cause project delay. Burgess et al. (2001) used BN to develop a software maintenance process, in which a software change requested in the next software release may far exceed the implementation resources available. The discussion above justifies our belief that IDs are suitable to model the risk factors involved in SDP and are capable of predicting the amount of training required by the staff.

### 3. ARCHITECTURE OF MAST

Based on research findings (Sharpa et al., 2001) and extensive interviews with 31 software professionals, we identified that due to the following risk factors, staff may be required to undergo training.

1. Lack of experience with project software (software involved in making the project was new to the employees).
2. Newly appointed staff (company was short of experienced staff and hence people who were fresh to the software industry were recruited).
3. Staff not well versed with the required quality standards (not aware of the standard of quality presently required, which may be due to increased priority and importance of present project or update in the company's CMM level).
4. Lack of experience with project environment (staff may not be familiar with the terminology or the processes involved in the project. For example, an employee involved may not be familiar with banking domain).

MaST used ID to calculate the percentage of staff training required and cost of conducting this training. The risk factors mentioned above were represented by chance nodes i.e. circles/ovals and 'Staff Training' which was dependent on the probability of occurrence of these risk factors was also represented by a chance node as shown in Figure 2. Because the above four risk factors have an impact on staff training, arcs are drawn with the tails on these risk factors and the heads toward 'Staff Training' to indicate the cause-effect relationship as discussed in Section 1.2. In other words, staff training node was driven by these four risk factors. Conducting staff training has a cost and was represented by utility node (diamond shaped) as shown in Figure 2. This cost was dependent on amount of staff training required. So, an arc was drawn from node 'Staff Training' to 'Cost'.

### 3.1 Tools used in the Implementation of MaST

To simplify the implementation phase, *Structural Modeling, Inference, and Learning Engine (SMILE)* (GeNIe: Web Address) have been used. SMILE is developed by Decision Systems Laboratory (DSL). SMILE is a platform independent library built in C++ classes for reasoning in BNs and IDs, which are a type of graphical probabilistic models. jSMILE is a Java Native Interface (JNI) library that enables the access to SMILE and SMILEXML C++ libraries from within Java applications. The current application is developed using NETBEANS (Java: Web Address), (an IDE for java) to develop the interface and jSMILE to make the ID. By using NETBEANS (for IDE) and jSMILE (to construct ID), MaST model was automated and included the features discussed below.

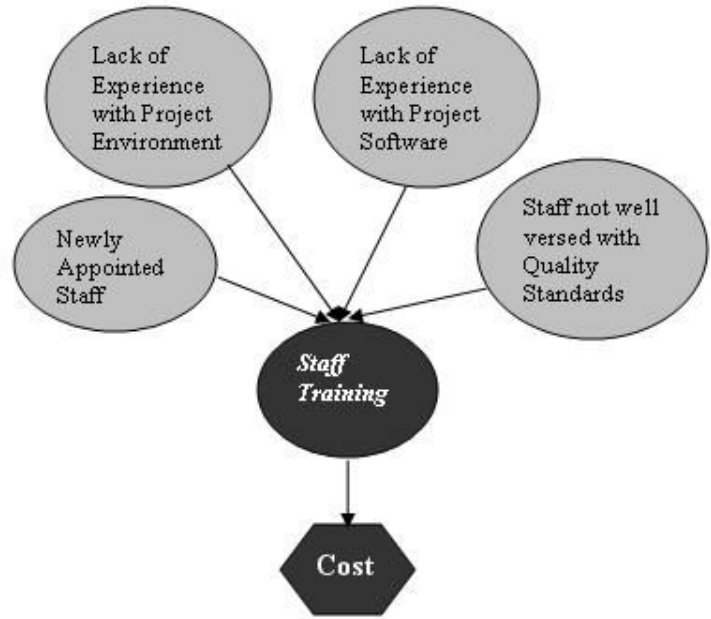

**Fig. 2:** Basic architecture of ID to calculate staff training and its cost

### 3.2 Functional Scales for MaST

For each node (of ID) of MaST, a measurement scale, i.e. a categorization such as probable, possible, yes, no, etc. of possible outcomes, was required. This categorization was required to be as close as possible to the way the management conducts assessment in that organization, so that MaST fits well into the organization. By expert interviews and experiments, the best scale for MaST chance nodes (mentioned in the beginning of Section 3) was found to be "Probable", "Possible", or "Remote". "Probable" if the risk was found to occur very frequently, "Possible" if found to occur less frequently, and "Remote" if found to occur rarely in the staff allocated to the present SDP. Similarly, for 'Staff Training' it was deduced to be "Yes", "No".

Prior probabilities, also called initial probabilities, were allocated to these outcomes by the project managers, on the basis of staff involved and the type of project under consideration. Let us assume that the managers found each of the three possible outcomes of each risk factor to be equally probable. So, the value of 1/3 = 0.33333 was assigned to each one of these. Similarly, prior probabilities for 'Staff Training' with both outcomes equally probable are 1/2 = 0.5 as shown in Figure 3.

The value node of an ID gives its outcome. In this case, the value node gives the cost of conducting the staff training. The arc into value node (from parent node 'Staff Training') indicated the event that affected the value of outcome. Cost node was accompanied by a value or utility table which reflects the cost of conducting the staff training when probability of 'Staff Training' was the maximum i.e. 1. This value was assumed to be $100,000 (experimental value) in this case, as depicted in Figure 3. To perform calculations, the categories of the outcomes of risk factors {Probable,

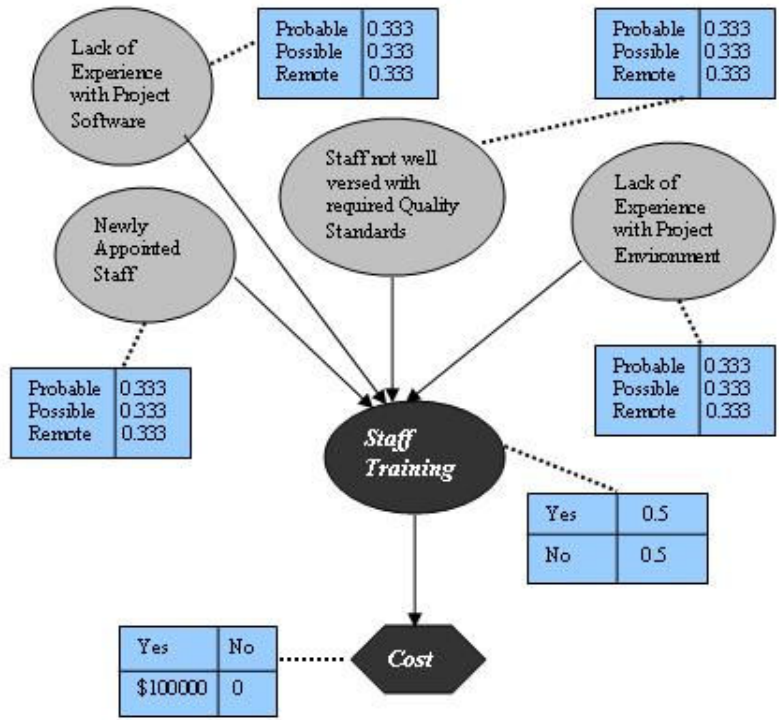

**Fig .3:** ID with probability tables for calculating staff training and its cost

Possible, Remote} were required to be mapped to some numeric values. The default for this was assumed to be {0.99999, 0.5, 0.00001}, meaning that if a particular risk factor's occurrence was highly probable, then its value would be 0.99999 (very high); if moderate, then it would be 0.5 (medium), and if significantly less, then it would be set to 0.00001 (very low).

All the risk factors do not have the same impact on staff training. So, after identifying the probability of risk, it was required to identify the risk severity level (i.e. loss that could occur due to a particular risk factor). The default scale for this impact was 0-10 (0 for the risk with negligible impact and 10 for the risk with catastrophic impact). The impact of the risk factor was calculated on the basis of experience.

**Fig. 4:** Interface to input impact of risk factors

**Fig. 5:** Interface to set the evidences/probabilities for risk factors

### 3.3 Implementation of MaST

As shown in Figure 4, project managers (users) can enter the impacts of all the four risk factors in the range 0-10. After accepting inputs from the user, MaST asked for another set of inputs, as shown in Figure 5. By using this interface, users can set the probability/likelihood of occurrence of the risk factors depending on the staff allocated to the project. On the other hand, in the background, MaST will generate

275

the ID (.xdsl file) that could be viewed by using GeNIe (see Figure 3).Worth mentioning is that GeNIe is a development environment (developed at the Decision Systems Laboratory) for building graphical decision-theoretic models.

To generate ID, conditional probability tables (CPT) were generated for the node 'Staff Training'. This table contains probabilistic information that relates conditional events, i.e., the four risk factors mentioned in Section 3. This CPT was calculated by MaST on multiplying probability of occurrence of each risk factor (r) and its impact or severity level (e) as entered in Figures 4 and 5. This was called risk exposure (re), see equation (1), because it calculates the exposure of system under consideration to overall risk of damage.

**re = r*e;**                                                         (1)

For each entry in CPT of "Staff Training"

(i) Identify the risk exposures (obtained using Eq. 1) of each risk factor for all possible outcomes and on the basis of impacts entered in Fig. 4.
(ii) Add contributory impacts of all of these factors to calculate overall exposure.

Let the risk exposure of any risk factor be *re* and the overall risk exposure or impact be *I*. The function *generateCPT* (provided later on) calculated the conditional probability for staff training corresponding to each possible outcome of the parent nodes (i.e., risk factors) and place them in an array *cpt* which was then populated in the actual node through jSMILE.

*Function generateCPT is*
*Input:*               *Possible values of all the four risk factors (Probable (Pr), Possible (Po), Remote (Re))*
*Risk factors impact weights w1, w2, w3, w4*
*Output:*            *CPT for Staff Training node (an array cpt[])*

**Algorithm**
*For each possible combination of outcomes of the four risk factors, calculate risk exposure (re) by multiplying each risk factor's possible outcome (Probable, Possible, Remote) with its corresponding weight*

```
    For all risk exposures (re)
            If re>=1 and <=2 add 0.5 to I
            If re>2 add <=3 add 1.0 to I
            If re>3 add <=4 add 1.5 to I
            If re>4 and <=5 add 2.0 to I
            If re>5 and <=6 add 2.5 to I
            if re>6 and <=7 add 3.0 to I
            If re>7 and <=8 add 3.5 to I
            If re>8 and re<=9 add 4.0 to I
            If re>9 and re<=10 add 4.5 to I
            Else add=0;
    End for
I=I/10
if (I>1)
    set I to 1 to show greatest impact, as the probability cannot be greater than 1
end if
put I at an appropriate position in cpt
end for
end function
```

**Explanation of the function *generateCPT***

The function 'add' was the contribution of a particular risk factor to the overall requirement of "Staff Training". These values were obtained by feedback from 31 software professionals and by conducting number of experiments. If the risk exposure *re* for a particular risk factor came to be less than 1.0, this meant that the risk had a negligible impact on staff training. In other words, the risk produced negligible loss and thus can be ignored. If *re* happened to be between 1.0 and 2.0, then it had a small impact and as a result, 0.5 was added to the overall impact as its contribution. Similarly, if *re* came to be between 2.0 and 3.0, then 1.0 was added as its contribution, and so on, meaning that as the risk exposure increased, the contribution of the risk of the risk factors to the requirement of staff training also increased. The overall risk exposure was divided by 10 to obtain the result in terms of the probability scale 0-1. Even if *I* happened to be greater than 1, *I* was reset to 1 as shown in the function *generateCPT*.

**TABLE 1**

Calculation of single conditional probability

| Risk Factor | Probability | Outcome (Value) | Impact | Risk Exposure (Value* Impact) | Risk Factor Contribution |
|---|---|---|---|---|---|
| Lack of experience with project software | Possible | 0.5 | 6 | 0.5*6=3.0 | 1.0 |
| Newly Appointed Staff | Remote | 0.00001 | 9 | 0.00001*9=0.00009 | 0 |
| Lack of experience with project environment | Probable | 0.99999 | 4 | 0.99999*4=3.99996 | 1.5 |
| Lack of Required Quality | Possible | 0.5 | 2 | 0.5*2=1 | 0.5 |

For instance, if the impacts for the risk factors were as shown in Figure 4 and probability of occurrence of the risk factors were as entered in Figure 5, then Table 1 shows how a single conditional probability value was calculated. According to the values entered in Table 1, "Lack of experience with project software" contributed 1.0 as its risk exposure (3.0) was between 2 an 3 ,"Newly Appointed Staff" did not contribute as its risk exposure was very low and had a negligible impact, "Lack of experience with project environment" contributed 1.5 as its risk exposure (3.99996) was between 3 and 4 and "Staff not Well Versed with Quality Standards" contributed 0.5 as its risk exposure (1) is between 1 and 2.

According to the function *generateCPT,* contributions from all the risk exposures are to be added. The overall impact of all risk factors I = 1.0 + 1.5 + 0.5 = 3.0.

Actual Conditional Probability (P) = I/10 = 0.3.

Table 2 was the partial CPT indicating the probability of the outcome ("Staff Training") being Yes or No for the four parent risk factors

As mentioned earlier, by pressing the 'submit' button shown in Figure 4, an interface shown in Figure 5 appears. The user entered the evidence (i.e., possible value of the risk factor) and by pressing the 'Inference' button, MaST accepted the evidence. MaST used the ID and made inference for the percentage of required staff training as

**TABLE 2**

Partial conditional probability table for staff training

| Lack of experience with project software | | | | | | | | Possible | | | | | | | | | | | | | | | | |
|---|---|---|---|---|---|---|---|---|---|---|---|---|---|---|---|---|---|---|---|---|---|---|---|---|
| Staff experience shortage | | | | | | | | Possible | | | | | | | | Remote | | | | | | | | |
| Lack of experience with project environment | | Probable | | | | Possible | | | | Remote | | | | Probable | | | | Possible | | | | Remote | | |
| Lack of Required Quality | | Pr | Po | Re | Pr | Po | Re | Pr | Po | Re | Pr | Po | Re | Pr | Po | Re | Pr | Po | Re | Pr | Po | Re | Pr | Po | Re |
| Staff Training | YES | 0.5 | 0.5 | 0.45 | 0.4 | 0.4 | 0.35 | 0.35 | 0.35 | 0.3 | 0.3 | *0.3* | 0.25 | 0.2 | 0.2 | 0.15 | 0.15 | 0.15 | 0.1 |
| | NO | 0.5 | 0.5 | 0.55 | 0.6 | 0.6 | 0.65 | 0.65 | 0.65 | 0.7 | 0.7 | 0.7 | 0.75 | 0.8 | 0.8 | 0.85 | 0.85 | 0.85 | 0.9 |

267

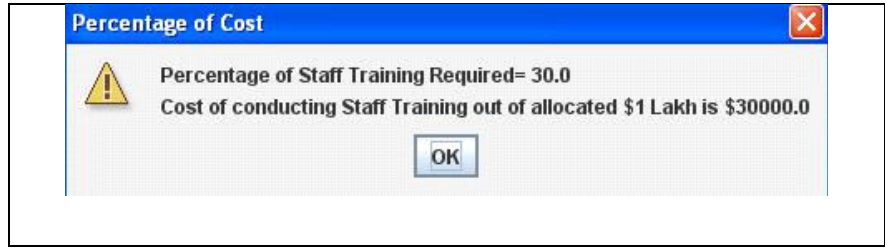

**Fig. 6 :** Probability of staff training and its cost: a sample output

well as its estimated cost. A message appeared as shown in Figure 6. This message showed the percentage of required staff training on basis of evidences set along with the amount of money needed to be spent on that staff training. For example, the percentage of the required staff training will be P*100 = 0.3*100 = 30%. The 'Clear Evidence' button in Figure 5 reset the evidence for a particular risk factor to null.

The output from the utility node (see Section 1.2 for details) is a value called the Expected Utility (EU). This value was computed using the utility Table 3.

**TABLE 3**

Utility Table for Staff Training

| Staff Training | Yes | No |
|---|---|---|
| Value (U) | $100000 | 0 |

This utility table represented the fact that if the value of "Staff Training" was 'Yes', then the expected utility was U, which was the cost of managing the training as per the requirement. If the value of "Staff Training" was 'No', then the expected utility was 0. Because the "Cost" was influenced by staff training, the outcome was dependent on the value of staff training, i.e., the cost of conducting the staff training was influenced by the amount of staff training required. If the probability of staff training happened to be 1.0, then the required cost of conducting the staff training would be $100000. So, for the case discussed earlier, the cost of conducting the staff training will be $100000 (value corresponding to staff training 'Yes') * 0.3 (staff training probability that was calculated) = $30,000.

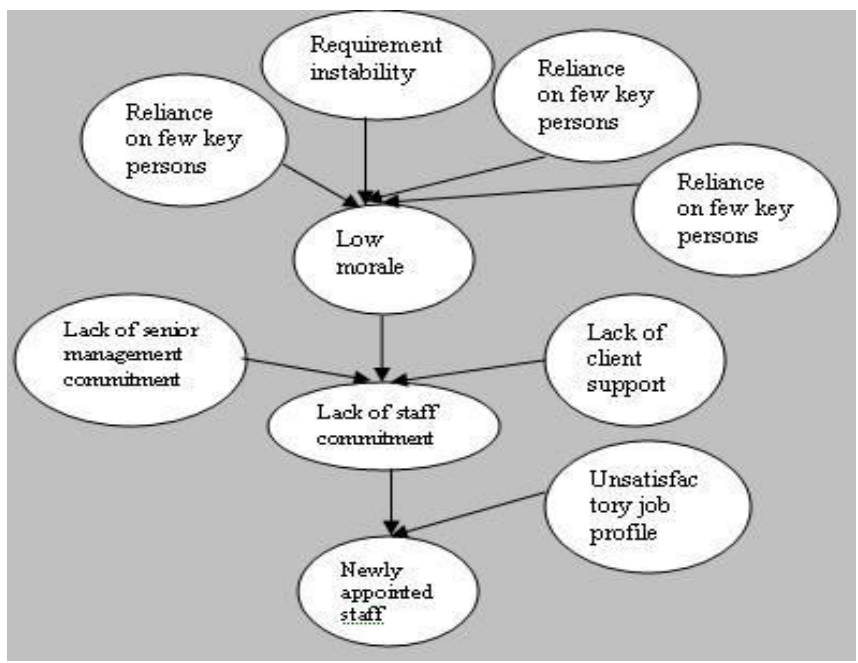

**Fig. 7:** Enhanced ID

### 4. DISCUSSION AND FUTURE ENHANCEMENTS

The success of software development process is measured by its cost, schedule, and performance (Conrow et al., 1997). So, after identifying the impact of Staff training on the cost, MaST can be further enhanced to calculate the impact of staff training on the schedule, i.e., the schedule overrun that could occur due to this staff training) and the performance (i.e., improvement in the performance of the employees due to this staff training).

As shown in Figure 7, "Newly Appointed Staff" is one risk factor that influences staff training and is further influenced by other risk factors like:
- Lack of staff commitment: If the staff is not committed, they will leave the organization resulting in a shortage of experienced staff.

281

- Unsatisfied job profile: If an employee is not satisfied financially or by the kind of work allocated he/she will leave the organization which leads to the appointment of new staff.

Some more risk factors directly or indirectly impact staff training but they are undergoing testing. For instance, "Lack of Staff Commitment" (Houston et al., 2001) is found to be influenced by:

- Lack of Senior Management Commitment: If senior management is not committed, then employees working under them will have lesser commitment.
- Lack of Client Support: If a client is not supportive, then automatically employees involved in the project will have reduced interest and so commitment.
- Low morale: If the employee is morally down, then he will have reduced commitment to the work and to the project. Low morale could be due to:
    1. Lack of contact person: If a person on the client side is not competent, he will not be able to communicate properly with the employees under consideration.
    2. Reliance on a few key people: If there is a reliance on just a few people or an employee is ignored and allocated to a work that is not up to his competence, it will result in a reduction of his morale.
    3. Requirement instability: If a client is changing the requirements very frequently, employees will get frustrated and it reduces their morale.
    4. Incapable project management: If a project is not managed properly by the management, it also affects the morale of the employee.

## 5. CONCLUSION

This paper has presented a study on using an ID to help manage training of the staff indulged in a SDP. In general, the amount of staff's training is hard to predict. Most of the available management tools are unable to pre-calculate the actual cost of conducting such a training program. But in the tool described here, we used IDs to store probability distributions of staff training based on risk factors to calculate a more accurate probability of staff training, as well as the cost of conducting such training. So far, we have applied this tool to a couple of projects and the results were quite encouraging.

**ACKNOWLEDGMENTS**

We wish to acknowledge the efforts of Mr. Harjit Singh, Senior Consultant, Tata Consultancy Services, Delhi, India; Pavneet Kaur, Lead Member Technical Staff, Mentor Graphics, Noida, India; Kawalpreet Kaur, System Administrator, Oracle India Pvt. Ltd., Noida, India.

**REFERENCES**

Abouzakhar, N.S., Gani, A., Manson, G., Abuitbel, M., King, D. 2003. *Bayesian Learning Networks Approach to Cybercrime Detection*, Postgraduate Networking Conference, Liverpool, United Kingdom.

Boehm, B., Abts, C., Chulani, S. 2000. Software development cost estimation approaches—A survey, *Annals of Software Engineering*, **10**, 177-205.

Boehm, B.W. 1991. Software Risk Management: Principles and Practices, *IEEE Software*, **8**, 32-41.

Burgess, C.J., Dattani1, I., Hughes, G., May, J.H., Rees, K. 2001.Using Influence Diagrams to Aid the Management of Software Change, *Requirements Engineering*,**6**, 173-82.

Charette, R.N. 1989. *Software Engineering Risk Analysis and Management*, New York, McGraw-Hill.

Chee, C.L., Vij, V., Ramamoorthy, C.V. 1995. Using influence diagrams for software risk analysis, *Seventh International Conference on Tools with Artificial Intelligence*, USA, IEEE,128-31.

Conrow, E.H., Shishido, P.S. 1997. Implementing risk management on software intensive projects, *IEEE Software*, **14**, 83-89.

Fan, C. Yu, Y. 2004. BBN-based software project risk management, *The Journal of Systems and Software, 73*, 193-203.

GeNIe: http://genie.sis.pitt.edu. Accessed 27 Feb 2008.

Houston, D.X., Mackulak, G.T., Collofello, J.S. 2001. Stochastic simulation of risk factor potential effects for software development risk management, *Journal of Systems and Software*, **59**, 247-57.

Hui, A.K., Liu, D.B. 2004. A Bayesian belief network model and tool to evaluate risk and impact in software development projects, *Annual Symposium-RAMS Reliability and Maintainability*, 297-301.

Java: www.sun.java.com Accessed 27 Aug, 2009.

Jeet, K., Mago, V.K., Prasad, B., Minhas, R.S. 2008. MaST: A tool for aiding the staff training management by using influence diagrams, *International Conference on Software Engineering Theory and Practice (SETP-08)*, Orlando, Florida, USA, 156-64.

Mago, V.K., Syamala, M., Mehta, R. 2008. Decision making system based on Bayesian network for an agent diagnosing child care diseases, Heidelberg, Springer Berlin, *Lecture Notes in Computer Science*, 127-136.

McCabe, B. 2001. Belief networks for engineering applications, *International Journal of Technology Management*, **21**, 257-70.

McManus, J. 2004. R*isk Management in Software Development Projects*, Burlington, Elsevier

Melo, A.C., de, Sanchez, A.J. 2007. Software maintenance project delays prediction using Bayesian Networks, *Expert Systems with Applications*, **34**, 908-19.

Norman, F.E., Martin, N. 2000. Software metrics: roadmap, *International Conference on Software Engineering*, Ireland, ACM, 357-70.

Owens, D.K., Shachter, R.D., Nease, R.F. 1997. Representation and Analysis of Medical Decision Problems with Influence Diagrams, *Medical Decision Making*, 17, 241-262.

Park, K.S. and Kim, S.H. 1998, Artificial intelligence approaches to determination of CNC machining parameters in manufacturing: a review, *Artificial Intelligence in Engineering*, **12**, 127-34.

Risk Management: www.businessbeam.com Accessed 27 Aug 2009.

Sharpa, H., Woodmanb, M., Hovendenc, F. 2004,Tensions around the adoption and evolution of software quality management systems: a discourse analytic approach, *International Journal of Human-Computer Studies,* **61**, 219-236.

Uffe, K.B.**,** Anders, M.L. 2008. *Bayesian Networks and Influence Diagrams: A Guide to Construction and Analysis,* Germany, Springer.